\documentstyle[11pt]{article}

\makeatletter

\def\revtex@ver{3.0}
\def\revtex@date{10 Jan 93}
\def\revtex@org{AAS}
\def\revtex@jnl{AAS}
\def\revtex@genre{preprint}
\def\revtex@pageid{\xdef\@thefnmark{\null}
\@footnotetext{This \revtex@genre\space was prepared with the
		   \revtex@org\space \LaTeX\ macros v\revtex@ver.}}
\def\genre@MS{manuscript}
\def\genre@PP{preprint}
\ifx\revtex@genre\genre@PP
\ifnum\@ptsize<1
\fi
\fi
\def\ps@plaintop{\let\@mkboth\@gobbletwo
\def\@oddfoot{}\def\@oddhead{\rm\hfil--\space\thepage\space--\hfil}
\def\@evenfoot{}\let\@evenhead\@oddhead}
\ps@plaintop
\@rightskip=\z@ plus 4em\rightskip\@rightskip
\def\@tightleading{1.1}
\def\@doubleleading{1.6}
\def\baselinestretch{\@doubleleading}
\def\tighten{\def\baselinestretch{\@tightleading}}

\def\singlespace{\def\baselinestretch{\@tightleading}\normalsize}
\def\doublespace{\def\baselinestretch{\@doubleleading}\normalsize}
\def\sec@upcase#1{\relax{#1}}
\def\eqsecnum{
\@newctr{equation}[section]
\def\theequation{\hbox{\normalsize\arabic{section}-\arabic{equation}}}}
\tighten
\def\@journalname{The Astropolitical Journal}
\def\cpr@holder{American Astronomical Society}
\def\received#1{\gdef\@recvdate{#1}} \received{\relax}
\def\revised#1{\gdef\@revisedate{#1}} \revised{\relax}
\def\accepted#1{\gdef\@accptdate{#1}} \accepted{\relax}
\def\journalid#1#2{\gdef\@jourvol{#1}\gdef\@jourdate{#2}}
\def\articleid#1#2{\gdef\@startpage{#1}\gdef\@finishpage{#2}}
\def\paperid#1{\gdef\@paperid{#1}} \paperid{MS-0001-SAMP}
\def\ccc#1{\gdef\CCC@code{#1}} \ccc{000-00\$75.95-CDB}
\def\cpright#1#2{\@nameuse{cpr@#1} \gdef\cpr@year{#2}
\typeout{`#1' copyright \cpr@year.}}
\newcount\@cprtype \@cprtype=\@ne
\def\cpr@AAS{\@cprtype=1}
\def\cpr@PD{\@cprtype=2}
\def\cpr@Crown{\@cprtype=3}
\def\cpr@none{\@cprtype=4}
\def\cpr@ASP{\@cprtype=5}
\def\cpr@year{\number\year}
\def\@slug{\par\noindent
\ifcase\@cprtype
	\relax
\or
	Copyright \cpr@year\space by the \cpr@holder.
\or
	This article is in the public domain.
\or
	Crown copyright \cpr@year\space by the \cpr@holder.
\or
	No copyright is claimed for this article.
\or
	Copyright \cpr@year\space by the ASP.
\fi
\par\noindent
Manuscript number \@paperid.\par\noindent
\CCC@code
}
\def\lefthead#1{\gdef\@versohead{#1}} \lefthead{\relax}
\def\righthead#1{\gdef\@rectohead{#1}} \righthead{\relax}
\def\@runheads{\@tempcnta\c@page
\@whilenum \@tempcnta >0\do{
\vskip 3ex
\hbox to30pc{\small\expandafter\uppercase\expandafter{\@versohead}:
	\expandafter\uppercase\expandafter{\@rectohead}\hfil}
\advance\@tempcnta by\m@ne}
}
\def\slugcomment#1{\gdef\slug@comment{#1}} \slugcomment{}
\newdimen\@slugcmmntwidth \@slugcmmntwidth .67\textwidth
\long\def\@makeslugcmmnt{\ifx\slug@comment\@empty\relax\else
\setbox\@tempboxa\hbox{\slug@comment}
\ifdim \wd\@tempboxa >\@slugcmmntwidth
\hbox to\textwidth{\hss
	    \parbox\@slugcmmntwidth\slug@comment}
\else
\hbox to\textwidth{\hfil\box\@tempboxa}
\fi
\vskip 2ex
\fi}
\def\@rcvaccrule{\vrule\@width1.75in\@height0.5pt\@depth\z@}
\def\@dates{{Received}\space%
\if\@recvdate\relax\@rcvaccrule\else\@recvdate\fi;%
\hspace{1.5em}{accepted}\space%
\if\@accptdate\relax\@rcvaccrule\else\@accptdate\fi%
}
\def\sluginfo{{\center
\@dates

\endcenter}}

\def\abstract{
\begin{center}
{\bf{ABSTRACT}}
\end{center}
\quotation
}
\def\title#1{\@makeslugcmmnt{\center\large\bf{#1}\endcenter}
\thispagestyle{empty}}
\def\author#1{{\topsep\z@\center\normalsize#1\endcenter}}
\let\authoraddr=\@gobble
\def\affil#1{\vspace*{-2.5ex}{\topsep\z@\center#1\endcenter}}
\def\altaffilmark#1{$^{#1}$}
\def\altaffiltext#1#2{\footnotetext[#1]{#2}\stepcounter{footnote}}
\def\and{\vspace*{-0.5ex}{\topsep\z@\center and\endcenter}}
\def\@keywordtext{Subject headings}
\def\@keyworddelim{---}
\def\keywords#1{\vspace*{-.7ex}
\if@twocolumn\noindent{{\it\@keywordtext:\/}\space\@kwds{#1}}
\else{\quote{\it\@keywordtext:\/}\space\@kwds{#1}\endquote}
\fi}

\def\@kwds#1{#1\relax}
\skip\footins 4ex plus 1ex minus .5ex
\newif\if@firstsection \@firstsectiontrue
\def\section{\if@firstsection
\@firstsectionfalse\fi
\@startsection {section}{1}{\z@}
{5ex plus .5ex}{1ex plus .2ex}{\normalsize\bf}}
\def\subsection{\@startsection{subsection}{2}{\z@}
{5ex plus .5ex}{1ex plus .2ex}{\normalsize\bf}}
\def\subsubsection{\@startsection{subsubsection}{3}{\z@}
{5ex plus .5ex}{1ex plus .2ex}{\normalsize\it}}
\def\thesection{\@arabic{\c@section}.}
\def\thesubsection{\thesection\@arabic{\c@subsection}.}
\def\thesubsubsection{\thesubsection\@arabic{\c@subsubsection}.}
\def\theparagraph{\thesubsubsection\@arabic{\c@paragraph}:}
\def\acknowledgments{\vskip 3ex plus .8ex minus .4ex}

\def\@sect#1#2#3#4#5#6[#7]#8{\ifnum #2>\c@secnumdepth
\def\@svsec{}\else
\refstepcounter{#1}\edef\@svsec{\csname the#1\endcsname\hskip 1em }\fi
\@tempskipa #5\relax
\ifdim \@tempskipa>\z@
\begingroup \center#6\relax
\@hangfrom{\hskip #3\relax\@svsec}{\interlinepenalty \@M
	  \sec@upcase{#8}\par}%
\endcenter\endgroup
\csname #1mark\endcsname{#7}\addcontentsline
{toc}{#1}{\ifnum #2>\c@secnumdepth \else
\protect\numberline{\csname the#1\endcsname}\fi
#7}\else
\def\@svsechd{#6\hskip #3\@svsec \sec@upcase{#8}\csname #1mark\endcsname
{#7}\addcontentsline
{toc}{#1}{\ifnum #2>\c@secnumdepth \else
\protect\numberline{\csname the#1\endcsname}\fi
#7}}\fi
\@xsect{#5}}
\def\@ssect#1#2#3#4#5{\@tempskipa #3\relax
\ifdim \@tempskipa>\z@
\begingroup #4\center\@hangfrom{\hskip #1}{\interlinepenalty \@M
\sec@upcase{#5}\par}\endcenter\endgroup
\else \def\@svsechd{#4\hskip #1\relax \sec@upcase{#5}}\fi
\@xsect{#3}}
\def\appendix{\par
\setcounter{section}{0}
\setcounter{subsection}{0}
\setcounter{equation}{0}
\def\thesection{\Alph{section}.}
\def\theequation{\hbox{\normalsize\Alph{section}\arabic{equation}}}}
\newcounter{cureqno}
\newenvironment{mathletters}{\refstepcounter{equation}%
\setcounter{cureqno}{\value{equation}}%
\let\@curtheeqn\theequation%
\edef\cur@eqn{\csname theequation\endcsname}%
\def\theequation{\cur@eqn\alph{equation}}%
\setcounter{equation}{0}}%
{\let\theequation\@curtheeqn%
\setcounter{equation}{\value{cureqno}}}
\def\eqnum#1{\def\theequation{#1}\let\@currentlabel\theequation
\addtocounter{equation}{\m@ne}}
\def\references{\subsection*{REFERENCES}
\bgroup\parindent=\z@\parskip=\itemsep
\def\refpar{\par\hangindent=3em\hangafter=1}}
\def\endreferences{\refpar\egroup\revtex@pageid}
\def\thebibliography{\subsection*{REFERENCES}
\list{\null}{\leftmargin 3em\labelwidth\z@\labelsep\z@\itemindent -3em
\usecounter{enumi}}
\def\refpar{\relax}
\def\newblock{\hskip .11em plus .33em minus .07em}
\sloppy\clubpenalty4000\widowpenalty4000
\sfcode`\.=1000\relax}
\def\endthebibliography{\endlist\revtex@pageid}
\def\@biblabel#1{\relax}
\def\@cite#1#2{#1\if@tempswa , #2\fi}
\def\reference{\relax\refpar}

\def\@citex[#1]#2{\if@filesw\immediate\write\@auxout{\string\citation{#2}}\fi
\def\@citea{}\@cite{\@for\@citeb:=#2\do
{\@citea\def\@citea{,\penalty\@m\ }\@ifundefined
{b@\@citeb}{\@warning
{Citation `\@citeb' on page \thepage \space undefined}}%
{\csname b@\@citeb\endcsname}}}{#1}}
\def\tablenotemark#1{\rlap{$^{\rm #1}$}}
\newtoks\@temptokenb
\def\tblnote@list{}
\def\tablenotetext#1#2{
\@temptokena={\vspace{.5ex}{\noindent\llap{$^{#1}$}#2}\par}
\@temptokenb=\expandafter{\tblnote@list}
\xdef\tblnote@list{\the\@temptokenb\the\@temptokena}}
\def\spew@tblnotes{
\ifx\tblnote@list\@empty\relax
\else
\vspace{4.5ex}
\footnoterule
\vspace{.5ex}
{\footnotesize\tblnote@list}
\gdef\tblnote@list{}
\fi}
\def\endtable{\spew@tblnotes\end@float}
\@namedef{endtable*}{\spew@tblnotes\end@dblfloat}
\let\tableline=\hline
\long\def\@makecaption#1#2{\vskip 2ex\noindent #1 #2\par}
\def\tablenum#1{\def\thetable{#1}\let\@currentlabel\thetable
\addtocounter{table}{\m@ne}}
\def\figurenum#1{\def\thefigure{#1}\let\@currentlabel\thefigure
\addtocounter{figure}{\m@ne}}
\newbox\pt@box
\newdimen\pt@width
\newcount\pt@line
\newcount\pt@nlines
\newcount\pt@ncol
\def\colhead#1{\omit\hidewidth{#1}\hidewidth\global\advance\pt@ncol by\@ne}
\def\tablecaption#1{\gdef\pt@caption{#1}} \def\pt@caption{\relax}
\def\tablehead#1{\gdef\pt@head{\hline\hline\relax\\[-1.7ex]
#1\hskip\tabcolsep\\[.7ex]\hline\relax\\[-1.5ex]}} \def\pt@head{\relax}
\def\tabletail#1{\gdef\pt@tail{#1}} \def\pt@tail{\relax}
\def\tablewidth#1{\pt@width=#1} \pt@width\textwidth
\def\tableheadfrac#1{\gdef\pt@headfrac{#1}} \def\pt@headfrac{.1}
\def\pt@calcnlines{\@tempdima\pt@headfrac\textheight
\@tempdimb\textheight\advance\@tempdimb by-\@tempdima
\@tempdima\arraystretch\baselineskip
\divide\@tempdimb by\@tempdima
\global\pt@nlines\@tempdimb}
\def\pt@tabular{\hbox \bgroup $\let\@acol\@ptabacol
\let\@classz\@tabclassz
\let\@classiv\@tabclassiv \let\\\@tabularcr\@tabarray}
\def\@ptabacol{\edef\@preamble{\@preamble \hskip \tabcolsep\tabskip\fill}}
\def\fnum@ptable{Table \thetable}
\def\fnum@ptablecont{Table \thetable---{\rm Continued}}
\def\set@mkcaption{\long\def\@makecaption##1##2{
\center\rm##1.\quad##2\endcenter\vskip 2.5ex}}
\def\set@mkcaptioncont{\long\def\@makecaption##1##2{
\center\rm##1\endcenter\vskip 2.5ex}}
{\crcr\noalign{\vskip .7ex}\hline\endtabular%
\pt@width\wd\pt@box\box\pt@box\spew@ptblnotes%
\typeout{Table \thetable\space has been set to width \the\pt@width}%
\endcenter\end@float}
\def\startdata{\pt@line=0\pt@calcnlines%
\ifdim\pt@width>\z@\def\@halignto{to \pt@width}\else\def\@halignto{}\fi%
\let\fnum@table=\fnum@ptable\set@mkcaption%
\@float{table}\center\caption{\pt@caption}\leavevmode%
\setbox\pt@box=\pt@tabular{\pt@format}\pt@head}
\def\pt@nl{\global\advance\pt@line by\@ne%
\ifnum\pt@line=\pt@nlines%
\endtabular\box\pt@box
\endcenter\end@float\clearpage%
\addtocounter{table}{\m@ne}%
\let\fnum@table=\fnum@ptablecont\set@mkcaptioncont%
\@float{table}\center\caption{\pt@caption}\leavevmode%
\global\pt@ncol=0%
\setbox\pt@box=\pt@tabular{\pt@format}\pt@head%
\global\pt@line=0%
\else\\
\fi}
\let\nl=\pt@nl
\let\nextline=\pt@nl

\def\tablebreak{\pt@line\pt@nlines\advance\pt@line by\m@ne\pt@nl}
\def\cutinhead#1{\noalign{\vskip 1.5ex}
\hline\pt@nl\noalign{\vskip -2.0ex}
\multicolumn{\pt@ncol}{c}{#1}\pt@nl
\noalign{\vskip .8ex}
\hline\pt@nl\noalign{\vskip -2ex}}
\def\sidehead#1{\noalign{\vskip 1.5ex}
\multicolumn{\pt@ncol}{@{\hskip\z@}l}{#1}\pt@nl
\noalign{\vskip .5ex}}
\def\set@tblnotetext{\def\tablenotetext##1##2{{%
\@temptokena={\vspace{0ex}{%
\parbox{\pt@width}{\hskip1em$^{\rm ##1}$##2}\par}}%
\@temptokenb=\expandafter{\tblnote@list}
\xdef\tblnote@list{\the\@temptokenb\the\@temptokena}}}}
\def\spew@ptblnotes{
\ifx\tblnote@list\@empty\relax
\else
\par
\vspace{2ex}
{\tblnote@list}
\gdef\tblnote@list{}
\fi}
\def\tablerefs#1{\@temptokena={\vspace*{3ex}{%
\parbox{\pt@width}{\hskip1em\rm References. --- #1}\par}}%
\@temptokenb=\expandafter{\tblnote@list}
\xdef\tblnote@list{\the\@temptokenb\the\@temptokena}}
\def\tablecomments#1{\@temptokena={\vspace*{3ex}{%
\parbox{\pt@width}{\hskip1em\rm Note. --- #1}\par}}%
\@temptokenb=\expandafter{\tblnote@list}
\xdef\tblnote@list{\the\@temptokenb\the\@temptokena}}
\def\thefigure{\@arabic\c@figure}
\def\fnum@figure{{\rm Fig.\space\thefigure.---}}
\def\thetable{\@arabic\c@table}
\def\fnum@table{{\rm Table \thetable:}}
\def\fps@figure{bp}
\def\fps@table{bp}
\let\jnl@style=\rm
\def\ref@jnl#1{{\jnl@style#1}}
\def\aj{\ref@jnl{AJ}}
\def\araa{\ref@jnl{ARA\&A}}
\def\apj{\ref@jnl{ApJ}}
\def\apjl{\ref@jnl{ApJ}}
\def\apjs{\ref@jnl{ApJS}}
\def\ao{\ref@jnl{Appl.Optics}}
\def\apss{\ref@jnl{Ap\&SS}}
\def\aap{\ref@jnl{A\&A}}
\def\aapr{\ref@jnl{A\&A~Rev.}}
\def\aaps{\ref@jnl{A\&AS}}
\def\azh{\ref@jnl{AZh}}
\def\baas{\ref@jnl{BAAS}}
\def\jrasc{\ref@jnl{JRASC}}
\def\memras{\ref@jnl{MmRAS}}
\def\mnras{\ref@jnl{MNRAS}}
\def\pra{\ref@jnl{Phys.Rev.A}}
\def\prb{\ref@jnl{Phys.Rev.B}}
\def\prc{\ref@jnl{Phys.Rev.C}}
\def\prd{\ref@jnl{Phys.Rev.D}}
\def\prl{\ref@jnl{Phys.Rev.Lett}}
\def\pasp{\ref@jnl{PASP}}
\def\pasj{\ref@jnl{PASJ}}
\def\qjras{\ref@jnl{QJRAS}}
\def\skytel{\ref@jnl{S\&T}}
\def\solphys{\ref@jnl{Solar~Phys.}}
\def\sovast{\ref@jnl{Soviet~Ast.}}
\def\ssr{\ref@jnl{Space~Sci.Rev.}}
\def\zap{\ref@jnl{ZAp}}

\def\deg{\hbox{$^\circ$}}

\def\lesssim{\mathrel{\hbox{\rlap{\hbox{\lower4pt\hbox{$\sim$}}}\hbox{$<$}}}}
\def\gtrsim{\mathrel{\hbox{\rlap{\hbox{\lower4pt\hbox{$\sim$}}}\hbox{$>$}}}}

\def\ion#1#2{#1$\;${\small\rm\@Roman{#2}}\relax}

\newcount\lecurrentfam
\def\LaTeX{\lecurrentfam=\the\fam \leavevmode L\raise.42ex
\hbox{$\fam\lecurrentfam\scriptstyle\kern-.3em A$}\kern-.15em\TeX}
\def\sizrpt{
(\fontname\the\font): em=\the\fontdimen6\font, ex=\the\fontdimen5\font
\typeout{
(\fontname\the\font): em=\the\fontdimen6\font, ex=\the\fontdimen5\font
}}

\makeatother
\tighten
\def\avrg#1{\langle{#1}\rangle}

\begin{document}

\title{The Cosmic X-ray Background -- IRAS galaxy Correlation
and the Local X-ray Volume Emissivity}

\author{Takamitsu Miyaji\altaffilmark{1,2}, Ofer Lahav\altaffilmark{3},
Keith~Jahoda\altaffilmark{1}, and Elihu~Boldt\altaffilmark{1}}
\begin{center}
 Accepted for publication in the ASTROPHYSICAL JOURNAL
\end{center}

\altaffiltext{1}{Laboratory for High Energy Astrophysics, Code 666,
NASA/Goddard Space Flight Center, Greenbelt, MD 20771, e-mail:
[miyaji,jahoda,boldt]@lheavx.gsfc.nasa.gov, mailing address for TM}
\altaffiltext{2}{Department of Astronomy, University of Maryland}
\altaffiltext{3}{Institute of Astronomy, Madingley Road, Cambridge,
 CB3 0HA, England,\\ e-mail:lahav@mail.ast.cam.ac.uk}

\begin{abstract}

 We have cross-correlated the galaxies from the IRAS 2 Jy redshift survey
sample and the 0.7 Jy projected  sample with the all-sky cosmic X-ray
background (CXB) map obtained from the HEAO-1 A2
experiment. We have detected a significant correlation
signal between surface density of IRAS galaxies and the X-ray background
intensity, with
$W_{xg}=\frac{\avrg{\delta I \delta N}}{\avrg{I}\avrg{N}}$
of several times $10^{-3}$.
While this correlation signal has a significant implication for the
contribution of the local universe to the hard ($E> 2\,keV$)
X-ray background, its interpretation is model dependent.

We have developed a formulation to model the cross-correlation
between CXB surface brightness and galaxy counts. This includes the
effects of source clustering and the X-ray -- far infrared luminosity
correlation. Using an X-ray flux limited sample of AGNs,
which has IRAS 60 $\mu m$ measurements, we have estimated the contribution
of the AGN component to the observed CXB -- IRAS galaxy count correlations
in order to see whether there is an excess component,
i.e. contribution from low X-ray luminosity sources. We have applied both
the analytical approach and Monte-Carlo simulations for the estimations.

 Our estimate of the local
X-ray volume emissivity in the 2 -- 10 keV band is
$\rho_{x}\approx (4.3 \pm 1.2)\times 10^{38}\,h_{50}\,
erg\, s^{-1}\, Mpc^{-3}$, consistent with the value expected from the
luminosity function of AGNs alone. This sets a limit to the local
volume emissivity from lower-luminosity sources (e.g. star-forming
galaxies, liners) to $\rho_{x}\lesssim 2 \times 10^{38}\,h_{50}\,
erg\, s^{-1}\, Mpc^{-3}$.

\end{abstract}


\section{Introduction}

  The origin of the Cosmic X-ray Background (CXB) is still an open question
(see review by Fabian \& Barcons 1992). The current popular idea of the
origin of the CXB is that it is a superposition of unresolved sources.
While much of the soft ($E<2 keV$) CXB  is resolved
into quasars by the ROSAT PSPC (Hasinger, Schmidt, \& Tr\"umper 1991;
Shanks et al.\ 1991; Hasinger et al.\ 1993), the discrepancy between the
$Log\, N - Log\,S$ relations of hard ($E\approx 2-10 keV$) and soft
bands ($E<2 keV$) may indicate that those sources are not the only
important contributor in the hard band.

 One important quantity which can be determined observationally is the
X-ray volume emissivity in the local universe.
This would provide a
constraint for models of the CXB with evolving populations of sources.
A lower limit to the quantity can be set from the X-ray luminosity function
(XLF) of resolved sources.
 The 2 -- 10 keV AGN XLF by Piccinotti et al. (1982) showed that it is
well approximated by a power-law function. The uncertainty on the
lower luminosity limit of the Piccinotti et al. XLF causes a large uncertainty
of the local volume emissivity from AGNs.
The AGN XLF is observed to flatten below $L_{x,44} \approx 0.05$
( where and hereafter $L_{x,44}$ refers to the 2 -- 10 keV X-ray luminosity
in units of $10^{44}\,h_{50}^{-2} erg\,s^{-1}$), which is the lower luminosity
limit of the XLF given by Piccinotti et al.\ (1982), by using a flux-limited
sample of AGNs (Grossan 1992, hereafter G92) extracted from the
{\em HEAO-1 MC-LASS Catalog of Identified, Hard X-ray Sources} (Remillard,
1994), which exploits the data from HEAO 1 A1/A3 experiments. This gives
$\rho_{x,38}\sim 4$ (where and hereafter, $\rho_{x,38}$ refers to the volume
emissivity in units of
$10^{38}\, h_{50} erg\, s^{-1}\, Mpc^{-3}$ in the 2 - 10 keV band unless
otherwise noted) for
$L_{x,44\, min} \sim 0.01$ (G92).

Also using the data from ROSAT and the
Einstein Observatory, Boyle et al. (1993) have determined the local
AGN luminosity function for
$L_{x,44}(0.3-3.5\, keV)>0.01$; this yields
$\rho_{x,38}(0.3-3.5\,keV) =1.7$. The corresponding volume emissivity implied
for the 2 -- 10 keV band has been estimated by Leiter \& Boldt (1994) based on
the unified model in which the number of Seyfert 2's is 2.3 times that of
Seyfert 1's (Huchra \& Burg 1992), the Seyfert 1 X-ray spectrum is the
relatively
steep one assumed by Boyle et al. (1993), and the average Seyfert 2
spectrum exhibits
absorption by $N_H \sim 5\times 10^{22} cm^{-2}$; this gives
$\rho_{x,38}(2-10\,keV)\sim 2.3$. For the G92 determined HEAO-1 based volume
emissivity (2-10 keV) to be consistent with the ROSAT value obtained by
Boyle et al. (1993) in the 0.3 - 3.5 keV band would require that the
number of Seyfert 2's be somewhat more than 4 times that of Seyfert 1's.

Extrapolating the volume emissivity implied by these resolved
sources to $z\sim 5$ explains 10 -- 30\% of
the 2 -- 10 keV CXB intensity without evolution.
 If there are a large number of low luminosity sources
($L_{x,44}\leq 0.01$, e.g. star-forming galaxies suggested by
Griffiths \& Padovani [1990] and/or liners), these would not appear in
the luminosity function, but would contribute to the CXB. Thus
an observational constraint for the local volume emissivity from these
sources gives an  important constraint to the models of the CXB.
%
Since galaxies are known to be clustered, one way to estimate the local
volume emissivity from low-luminosity sources is to search for the auto-
correlation property of the CXB at a scale of degrees. An analysis using
this method by Danese et al. (1993) set the upper limit to the local
X-ray volume emissivity
clustered like normal galaxies to $\rho_{x,38}\lesssim 6$.

  Another  method to probe the volume emissivity due to these low luminosity
sources is to cross-correlate the CXB surface brightness with known catalogs
of galaxies. The first such attempt was made by Turner \& Geller (1980)
who set an upper limit to the fraction of the CXB correlated with
nearby galaxies. Persic et al. (1989) took a similar but slightly different
approach. They looked for a surface brightness enhancement at positions
of various classes of sources and  set an upper limit to the constribution
of low-luminosity sources to CXB.
Jahoda et al. (1991; 1992, hereafter JLMB91, JLMB92 respectively
or JLMB collectively) have used the HEAO-1 A2 all-sky
X-ray map to cross-correlate with the galaxies in the UGC and ESO catalogs
and found a correlation signal. Based on this
correlation signal and effective depths of UGC and ESO catalogs, they have
derived a local X-ray volume emissivity in the 2--10 keV band of
$ \rho_{x,38}=(12.5 \pm 7.0)$~\footnote{The value was initially given
incorrectly in JLMB91 (Erratum: JLMB92). We show the corrected value here.}.
Their calculation, however, was
based on the assumptions that the effect of source clustering can be
neglected and that the radial selection functions of cataloged
galaxies and the
associated X-ray sources are identical. Lahav et al.\ (1993) estimated the
clustering effect and concluded that the effect is not at all
negligible and that one thereby overestimates the volume emissivity by
a large factor.

 In this work, we have used two complete flux limited samples of
galaxies from the IRAS point source catalog,
i.e. the IRAS 2~Jy redshift survey (Strauss et al.\ 1990)
and the 0.7 Jy (Meurs \& Harmon 1988) projeted samples to correlate with
the all-sky
CXB surface brightness map from the HEAO-1 A2 experiment. The 2~Jy sample is
particularly useful because it has a redshift for each galaxy and thus we can
calculate correlation coefficients for a few redshift-selected subsets of the
sample. In \S~\ref{sec:data}, we briefly summarize the data we have used and
also explain the calculation of the correlation coefficients.
In \S~\ref{sec:ana} and \ref{sec:appl}, we develop analytical formulations
of the correlation and the application of the formulations to this particular
problem. This includes a full treatment of the effects of source clustering
and the X-ray -- 60 $\mu m$ luminosity correlation. Some details of the
derivations are shown in appendices A - B. 
The Monte-Carlo simulations are explained in \S~\ref{sec:sim} and
and the simulated models are compared with the observations as well as
analytical calculations in \S~\ref{sec:comp}. The results are discussed
in \S~\ref{sec:dis}

\section{The Data and Correlation Analysis}
\subsection{The Data}
\label{sec:data}
The X-ray data are from all-sky HEAO-1 A2 survey (Rothschild et al.\ 1979;
Boldt 1987). The X-ray map has been constructed with the data from MED and
HED \# 3 of the A2 experiment (Jahoda \& Mushotzky 1989; JLMB91) accumulated
over the 180 day period beginning on day 322 of 1977. The
period corresponds to one complete scan over the entire sky. The
combination had sensitivity from 2 to 60 keV and quantum efficiency over 50\%
roughly between 3 and 17 keV. The all-sky X-ray map constructed from the
data observed with the $1\deg .5 \times 3\deg .0$ FWHM collimators are used
for the analysis. The point spread function (PSF) of the observation through
these collimators can be well represented by
$B_{psf} = max( 1-\frac{|x|}{3\deg} ,0 ) max( 1-\frac{|y|}{1\deg .5}, 0)$,
where $y$ is the
coordinate along the scan path of the survey and $x$ is along the
axis perpendicular to it (Shafer 1983). For
this combination, one count per second corresponds to a 2 -- 10 keV flux of
$\sim 2.1\times 10^{-11}\,erg\,s^{-1}\,cm^{-2}$ for the 40 keV thermal
bremsstrahlung emission. The conversion differs only by a few percent between
a power-law spectrum with the energy index of 0.65 and a 40 keV
bremsstrahlung (Shafer 1983).

The map has been corrected for the Compton-Getting Effect (e.g. Boldt 1987)
which is the all-sky dipole distortion due to the sun's peculiar motion
at 370 $km\,s^{-1}$ toward  $(l^{II},b^{II})=(264\deg .4, 48\deg .4)$
inferred from the COBE DMR measurement of the cosmic microwave background
dipole anisotropy (Kogut et al. 1993).
A slight ($\sim 3\%$) sensitivity change over the 180 day period has been
recognized. The map has been corrected for this effect by linearly modeling
the sensitivity change.

The IRAS 2~Jy sample complete with redshift values
(e.g. Strauss et al.\ 1990) and a deeper (0.7 Jy) sample of color selected
IRAS galaxies (Meurs \& Harmon, 1989) have been
correlated with the CXB intensity. For the 2~Jy sample, we have limited
our analysis to those galaxies with radial velocities between 500 and 8000
$km\,s^{-1}$, where
the selection function is well defined (Strauss et al. 1992a).
Although the 2 Jy
sample is shallow, it has an advantage of having redshift on each
object so the volume correlated with CXB is well defined.
For the 0.7 Jy sample, we have used the selection function based
on the luminosity function by Saunders et al. (1990). This sample
is more sensitive to the low-luminosity behavior of the luminosity function.

%
%
\subsection{The Correlation Analysis}

To avoid confusion with galactic sources, we have limited our
analysis to $|b|>20\deg $; regions around the Magellanic clouds are
excluded. We have also excluded regions within one beam
of sources in Piccinotti et al.\ (1982). But we have included nearby
AGNs in that sample which may contribute to the correlation signal,
i.e. AGNs with  $v<8000 km\,s^{-1}$
for the correlation with the 2 Jy sample and $z<0.1$ for the correlation
with the 0.7 Jy sample. These have been kept in the analysis
because calculating the correlations including them is more convenient
for the comparison with the formulations developed in the next section.
In this work, we are not
interested in the contributions from the X-ray emission of clusters
of galaxies. They are rare in number density and known to evolve negatively
with redshift (Edge et al. 1990); therefore their contribution to the CXB is
not very significant. Thus we excluded data from within one beam of clusters
in Edge et al. (1990). Since the depths of the
IRAS samples are shallow, the contribution from the X-ray clusters fainter
than the flux limit of Edge et al. to the correlation signal is expected to
be small.
The regions of the sky which is not covered by the IRAS samples
(see Strauss et al.\ [1990] for the 2Jy sample and Meurs \& Harmon [1989]
for the 0.7 Jy sample)are also excluded from the analysis.

The HEAO-1 A2 all-sky map is correlated with the galaxy surface density map
created by smearing the position of the selected galaxies
with the PSF of the A2 experiment. We have calculated the zero-lag
cross-correlation in $3\deg \times 3\deg$ square cells.
 The gridding has been made with a coordinate system which has one of
its poles at the galactic center and the zero-longitude great circle on the
galactic plane. Using this coordinate system, we have divided the sky into
latitude strips of $3\deg$ width and subdivided each latitude strip every
$3 [cos(b_c)]^{-1}$ degrees in longitude, where $b_c$ is the central latitude
of the strip. With this division process, cells are made nearly
$3\deg \times 3\deg$ square except near the coordinate poles, which are in
the galactic plane. We have accepted the
cells for analysis only if the excluded region
is no more than 20\% of the cell. In this case, the small excluded region is
assumed to have the mean X-ray surface brightness and galaxy surface density
as the rest of the cell.  This procedure leaves about
2300 cells for analysis, covering about a half of the whole sky.

As a statistical characterization of the zero-lag correlation, we have
calculated the correlation coefficient:
\begin{equation}
W_{xg}=\frac{N_{cells}\,\sum_i (I_i-\langle I \rangle )
(N_i-\langle N \rangle)}{(\sum_i I_i)(\sum_i N_i)}
\sim \frac{\avrg{\delta I\,\delta N}}{\avrg{I} \avrg{N}},
\end{equation}
where $I$ is the surface brightness of the X-ray background,
$N$ is the surface number density of the sample galaxies, and the summation is
over $N_{cells}$ cells. We have correlated the CXB with subsets
of IRAS galaxies selected by the redshift range and the supergalactic
latitude ($|SGB|$). The values of $\avrg{I}$ and $\avrg{N}$ are for the
subsets and not the global averages. The results of the correlation are
summarized in Table~\ref{tbl:corr} with the 1$\sigma$ errors estimated from
the bootstrap resampling of the correlated data (JLMB and references
therein). We measure stronger correlation signals inside
the supergalactic plane ($|SGB|<20\deg$) than outside of it. The enhancement
at low $|SGB|$ is dominated by low redshift IRAS galaxies
($V < 3500\,km\,s^{-1}$).
We analyze the all-sky average behavior of the correlation signals in
different redshift bins in the following section. For reference, we also
show the case where all the AGNs in Piccinotti et al. (1982) are also
excluded from analysis (fourth row in Table~\ref{tbl:corr}). This shows
that there still is a significant correlation between IRAS galaxies
and off-source X-ray sky. The exclusion procedure shown above, however,
would make the analysis in the following section complicated and thus we
model the correlation signal when resolved AGNs are included as described
above.

\section{Correlation and the X-ray Volume Emissivity}
\label{sec:corr}

\subsection{Analytical Formulations}
\label{sec:ana}
In this section, we develop formulations relating the
X-ray volume emissivity and the correlation
signal. These are  useful for investigating the dependence of
the correlation signal on various parameters.
In particular, we have included the effects of the source clustering
and the X-ray -- 60
$\mu m$ luminosity correlation (see also Lahav et al. 1993). The source
clustering causes an enhancement of the correlation signal compared with
the Poisson case for a fixed volume emissivity. The effect
is partially due to the enhanced fluctuations associated with
 the angular clustering
of the sample galaxies, some of which are X-ray sources. The effect is
also due to the X-ray sources which themselves
are not in the catalogued galaxies, but are clustered with the sample
galaxies. If X-ray and infrared luminosities are correlated, the Poisson
process should also be
enhanced compared with the case with no such correlation.

 We observe the galaxy counts and the X-ray intensities through effective
cells, where each is the convolution the instumental PSF
(point spread function)
and the square box profile of the cell. (Hereafter, the expression a 'square
cell' refers to a cell with a square profile as well as a square
projected shape.) We express the effective cell profile
$B_{ec}(\hat{R}-\hat{R_0})$ normalized at the center ($B_{ec}({\bf0})=1$),
where $\hat{R}$ and
$\hat{R_0}$ are the unit vectors of the current position and the cell center
respectively. We also define the effective cell solid angle
$\Omega _{ec} \equiv \int d\Omega B_{ec}(\hat{R}-\hat{R_0})$.
 As Lahav et al. (1993), we express the expected
correlation as a sum of the Poisson ($\hat{\eta _p}$) and the clustering
($\hat{\eta _c}$) terms:

\begin{equation}
W_{xg}\langle I \rangle \langle N \rangle
\equiv
\avrg{(N- \avrg{N} )(I-\avrg{I})}  = \hat{\eta _p} + \hat{\eta_c}
\label{eq:wxg}
\end{equation}
Note that here $I$ and $N$ are the X-ray intensity and the IRAS galaxy number
density per unit solid angle while the same symbols represent {\em per
beam} values in Lahav et al. (1993).

As detailed in appendices A \& B 
, the Poisson term
can be expressed as:
\begin{equation}
\hat{\eta}_p= \frac{\int d\Omega\,B_{ec}^2(\hat{R}-\hat{R_0})}
{4\pi \Omega_{ec}^2}\rho _{xp} R_p\equiv \frac{s_p}{4\pi \Omega_{sq}}
\rho _{xp} R_p,
\label{eq:etap}
\end{equation}
where the R.\ H.\ S.\ of the equation expresses
in terms of the square cell case with $\Omega_{sq}=a^2$ and a
factor $s_p$ explaining the PSF smearing effect for convenience.

 In the general case where $L_x$ may be correlated
with $L_{60}$, the Poisson term effective depth $R_p$ can be
expressed as:
\begin{mathletters}
\begin{equation}
R_{p} \equiv \int_{R_{min}}^{R_{max}} dR P_{x}(R),
\end{equation}
\begin{equation}
P_x(R) = \frac{\int_{4\pi R^2 f_{60,lim}}^{\infty} dL_{60}
\bar{L}_x(L_{60}) \Phi_{60}(L_{60})}{\int_0^{\infty}
dL_{60} \bar{L}_x(L_{60}) \Phi_{60}(L_{60})}\,,
\label{eq:pxr}
\end{equation}
\end{mathletters}
where $\Phi _{60}(L_{60})$ is the 60$\mu m$ luminosity function and
$\bar{L}_x(L_{60})$ is the mean X-ray luminosity {\em per IRAS source} of
a given 60$\mu m$ luminosity.
Note that if $\bar{L}_x$ is constant (no correlation
between $L_x$ and $L_{60}$), then $P_x(R) = P(R)$ (the usual selection
function).

Likewise, we express the clustering term. For the spatial correlation
function between X-ray sources and IRAS galaxies
$\xi_{xg}(r)$, the clustering term can be expressed as
(Lahav et al. 1993):
\begin{eqnarray}
\hat{\eta}_c =\frac{1}{\Omega_{ec}^2} \int d^3 {\bf R_1} \int d^3 {\bf R_2}
\;\frac{\rho_{xc}}{4\pi R_2^2}\avrg{n}P(R_1)
\;\; \xi_{xg}(|{\bf R_2} -{\bf R_1}|)
B_{ec}(\hat{R}_1-\hat{R}_0)\,B_{ec}(\hat{R}_2-\hat{R}_0) \nonumber \\
\equiv \frac{s_c}{4\pi \Omega _{sq}} \rho _{xc}R_c,
\label{eq:etac}
\end{eqnarray}
where $\avrg{n}$ is the mean number density of
the galaxies in the sample (above some minimum luminosity), $P(R)$ is
the selection function of the sample, $\hat{R}_1$ \& $\hat{R}_2$ are the
unit vectors towards ${\bf R_1}$ \& ${\bf R_2}$ respectively.
Here again, the effect of the instrumental PSF is expressed by the
factor $s_c$.
Assuming a power-law spatial correlation function
between IRAS galaxies and X-ray sources
$\xi _{xg} = (r/r_{0})^{-\gamma}$
, and also assuming that the scale of the clustering
is much smaller than the distances to the objects
($|{\bf R_1}-{\bf R_2}|\ll R_1,\, R_2$) (appendix~C), 
\begin{equation}
R_c \equiv \int_{R_{min}}^{R_{max}} dR P_c(R)
\approx \avrg{n} A_{\gamma} r_0^{\gamma} a^{3-\gamma}
\int_{R_{min}}^{R_{max}} dR R^{3-\gamma} P(R),
\label{eq:pcr2}
\end{equation}
where $A_{\gamma}$ is the geometrical factor for the $\Omega_{sq}=a^2$ square
cell depending on $\gamma$.
While the volume emissivity in the Poisson term
$\rho_{xp}$ is due to X-ray sources among the catalogued class of
galaxies (i.e ,those having IRAS luminosities above $L_{60,min}$),
 $\rho_{xc}$ is due
to X-ray soueces {\em clustered} with
those galaxies, which may or may not be among those galaxies (if all the
X-ray emitters responsible for the correlation are in the catalogued
{\em class} of galaxies, then $\rho _{xp} = \rho _{xc}$).

\subsection{Application to the HEAO 1 -- IRAS  Case} \label{sec:appl}

 We now apply the general formalism developed above to our problem.
In the case of the $3\deg \times 3\deg$ square cells
($\Omega_{sq}=9\,deg^2$) convolved with the A2 PSF we have used here,
$\Omega_{ec} \approx 12\,deg^2$, and $s_p\approx 0.42$. These values
are insensitive (within a few percent) to the alignment between the
square dividing cell and the scan path (i.e. orientation of the
PSF) of the A2 experiment.
 For $\gamma=1.8$, $s_c \approx 0.74$ and $A_{1.8} \approx  8.3$.
(If the instrumental PSF could be considered a delta function,
$\Omega_{sq}=\Omega_{ec}$ and $s_p = s_c = 1$.)

First, we consider the expected correlation signal from the population
of X-ray sources whose X-ray luminosity function is reasonably
known from the existing catalogs.
The MC-LASS catalog of X-ray sources
(Remillard 1994) include identifications of the
X-ray sources using the data from HEAO 1 A1 (LASS) and A3 (Modulation
Collimator) experiments. Grossan (G92) has made extensive studies
of a complete flux limited sample of AGNs extracted from the MC-LASS
catalog (flux limit: 0.95 $\mu Jy$ at 5 keV,
which is about a factor of two lower than that  of Piccinotti et al.
[1982]) (the LMA sample [LASS/MC identified sample of AGNs]). He
constructed the
luminosity function of the LMA AGNs giving $\rho_{x,38}=4.1_{-1.7}^{+1.3}$,
including AGNs down to $L_{x,44} \approx 0.01$. There are also
extensive measurements over the electromagnetic spectrum for
the LMA AGNs including redshifts and IRAS $60 \mu m$ fluxes (G92).
%
By comparing the observed correlations and the expected
correlations from those AGNs, we can estimate or set a limit
to the contribution of the lower-limunosity sources to the local
volume emissivity.

As shown in appendix~A, the numerator of the R.\ H.\ S.\
of Eq.~\ref{eq:pxr} ($=\rho_{xp}P_x(R)$) can be estimated using
a complete X-ray flux limited sample with redshifts and $60\mu m$ flux
measurements ($L_{x,i}\,, L_{60,i}$). Neglecting the clustering of the
sample sources, this is:
\begin{equation}
\int_{4\pi R^2 f_{60,lim}}^{L_{60,max}} dL_{60} \bar{L}_x(L_{60}) \Phi_{60}
(L_{60}) \approx \sum_{L_{60,i}\geq 4\pi r^2 f_{60,lim}}
\frac{L_{x,i}}{V_{max}(L_{x,i})},
\label{eq:pxr2}
\end{equation}
where $f_{60,lim}$ is the limiting flux of the correlating IRAS sample and
$V_{max}(L_x)$ is the volume of space where a galaxy with an X-ray luminosity
$L_x$ would be in the X-ray flux limited sample.

 As discussed in appendix~A, $P_x(R)$ estimated using
Eq.~\ref{eq:pxr2} only represents the contribution to the Poisson
term from the sources with the X-ray luminosities
covered by the sample used to
evaluate Eq.~\ref{eq:pxr2}. Another Poisson term should be added for
the contribution of lower luminosity objects.
 Thus it may be is convenient to divide the Poisson term into two parts:
\begin{equation}
W_{xg}\avrg{I} \avrg{N}
=\frac{1}{4\pi \Omega_{sq}} [s_p(\rho_{xpA}R_{pA} + \rho _{xpB}R_{pB})
 + s_c \rho_{xc}R_c],
\label{eq:wxgl}
\end{equation}
where $\rho _{xpA,38}= 4.1_{-1.7}^{+1.3}$ (G92) is the volume
emissivity of sources
with $L_{x,44}\geq 0.01$ (component [A]) and $\rho _{xpB}$ is the X-ray
volume emissivity
of the sources which are in the IRAS catalog (i.e.$L_{60}\geq L_{60,min}$),
with ($L_{x,44}<0.01$)(component [B]), for which we do not know the X-ray
luminosity function. If most of the X-ray sources emit far infrared radiation
also with $L_{60}\geq L_{60\,min}$, then $\rho_{xc}\sim \rho_{xpA} +
\rho_{xpB}$.

  Applying Eq.~\ref{eq:pxr2} to the 61 AGNs from the LMA sample with
radial velocities less than 20000 $km\,s^{-1}$ (Fig.~\ref{fig:lma}),
we have evaluated $P_{xA}(R)$
($P_x(R)$ for component [A]) and $R_{pA}$. Out of the 61 objects,
8 objects have only upper limits to the 60 $\mu m$ flux. For these 8 objects,
we have assigned 60~$\mu m$ luminosities
corresponding to a half of their 3$\sigma$ upper limit fluxes. We also
compared the resulting $R_{pA}$ by excluding the 8 objects and observed
that the effect of including/excluding these objects are much smaller than
other sources of errors.

 To evaluate the clustering term, we note that the spatial
correlation function $\xi_{xg}(r)$in Eq.~\ref{eq:etac} is the
correlation function between IRAS galaxies and X-ray sources.
An estimate of this correlation function from the available data
may be found by finding the cross-correlation function between
resolved X-ray sources and the IRAS galaxies. Since both the IRAS 2 Jy
sample and the LMA sample have redshifts, it is easy to calculate
the spatial correlation function between these.
We estimate $\xi_{xg}(r)$ by:
\begin{equation}
\xi_{xg}(r)=\frac{N_{pair}(r)}{\hat{N}_{pair}^R(r)} - 1,
\end{equation}
where $N_{pair}(r)$ is the number of X-ray AGN - IRAS pairs separated
by a distance $r$ and $\hat{N}_{pair}^R(r)$ is the ensemble average
of the number of pairs separated by the same distance between
randomized X-ray AGN and IRAS samples. The redshifts of the randomized
samples are drawn from the real samples but the sky coordinates are
randomized within the sampled region. This method of constructing
randomized samples compensates for the effects of the
radial selection functions and incomplete sky coverage of the
LMA and IRAS 2 Jy samples.
 The result is
shown in Fig.~\ref{fig:xi}. Fig.~\ref{fig:xi} shows that
the IRAS-LMA AGN correlation function is well represented by
a power-law form $\xi _{xg} = (r/r_{0})^{-\gamma}$ with the
$r_0 \sim 400\,km\,s^{-1}$ and $\gamma \sim 1.8$ except
about a factor of two deficit at $r\approx 100 - 200 \,km\,s^{-1}$.
Here and in the next subsection, we use the power-law spatial
correlation function with $r_0 \sim 400\,km\,s^{-1}$ and
$\gamma \sim 1.8$. This result is about a factor of two smaller than the
IRAS QDOT - IRAS selected AGN correlation function
(Georgantopoulos \& Shanks 1993). The origin of the
discrepancy is unclear.  Also it is somewhat steeper than
the IRAS autocorrelation function
($r_0\sim 400\,km\,s^{-1}$,$\gamma \sim 1.6$
; Lahav, Nemiroff, Piran 1990; Strauss et al. 1992a;
Saunders, Rowan-Robinson, \& Lawrence 1992). One source of systematc error
may be  that we have calculated $\xi_{xg}$ using the redshift instead of the
real space distance. Sensitivity of the results
to the assumed correlation function is discussed in \S~\ref{sec:dis}

To evaluate Eqs.~\ref{eq:etac} \& \ref{eq:pcr2}, we need the radial
selection function of the IRAS sample $P(R)$.  We used the Strauss et al.
(1992a) selection function for the 2 Jy sample. We have constructed the
radial selection function from the luminosity function by Saunders et al.
(1990) for the 0.7 Jy. The 0.7 Jy luminosity selection function has been
normalized to unity at $v = 500\,km\,s^{-1}$ and assumed to be unity
at nearer distances.  Contribution
of $v < 500\, km\,s^{-1}$ to $R_c$ for the 0.7 Jy sample is small
($\sim$ 2\%).
The lower and upper bounds of the integration to obtain
$R_c$ and $R_{pA}$ for the 0.7 Jy sample were set at 0 and 20000 $km s^{-1}$
respectively. The upper bound was determined to give the observed surface
number density for the given form of selection function. The calculated values
of $R_{pA}$ and $R_c$ are summarized later along with the
Monte-Carlo simulation results (\S~\ref{sec:comp}).


The selection functions multiplied by the smearing factors
$s_cP_c(R)$,and $s_pP_{xA}(R)$ evaluated
above are plotted in Fig.~\ref{fig:sel} along with $s_pP(R)$.
Fig.~\ref{fig:sel} shows relative importance of each term as a function
of redshift as compared to the purely Poisson no-luminosity correlation
case (JLMB)($s_pP(R)$ curve). Although Fig.~\ref{fig:lma} shows only
a weak correlation between $L_x$ and $L_{60}$ for the X-ray selected
sample, the increased fraction of the luminous X-ray sources with larger
$L_{60}$ makes $P_{xA}(R)$ much larger than $P(R)$ at large distances.
(Remember that $\bar{L}_x(L_{60})$ in Eq.~\ref{eq:pxr} is the mean X-ray
luminosity {\em per IRAS source} with $L_{60}$.)
  The peaked feature of $s_cP_c(R)$ of the 0.7 Jy
(Fig.~\ref{fig:sel}(b)) can be understood as
follows. The clustering selection function $P_c(R)$ includes the product of
$P(R)$ and the volume integration of $\xi _{xg}$. The former is a rapidly
decreasing function of $R$, while the latter increases with distance as
an effective cell covers more of the clustered sources.

 This analytical approach is useful in looking at the behavior
of Poisson and clustering terms, investigating the sensitivity of the
results to parameters, and understanding the principles of the
surface brightness - number count correlations. However,
the depth of the IRAS samples are such that the approximation
used in Eq.~\ref{eq:pcr2} is not accurate and an elaborate Monte-Carlo
integration is needed to evalute the clustering term without
this approximation. Also integrating the power-law correlation
function to infinity may cause systematic errors. Also statistical
uncertainties due to the finite
number of sources contributing to the correlation are hard to estimate
with the analytical approach. Thus we use the Monte-Carlo simulations
to compare the models with the data in the next section.

\subsection {Monte-Carlo Simulations}
\label{sec:sim}
We have made Monte-Carlo simulations in order to verify our
analytical formulations and estimate the uncertainties of the correlations.
The IRAS particles are drawn from N-body CDM simulation particles
provided to us by White (1993),
which is characterized by a spatial correlation function with a power-law
index $\gamma \sim 1.8$ and a correlation length of 0.066 of the size
of the box. The cubic box includes 9040 particles with a periodic
boundary condition. We have scaled the volume  to the correlation
length of $r_0 = 400\, km\, s^{-1}$ and assigned each CDM particle
$60 \mu m$ and X-ray  luminosities ($L_{60}$ \& $L_x$). Fig.~\ref{fig:xi}
shows the spatial correlation function of the scaled CDM particles as
open triangles compared with the power-law form used for the analytical
calculations and the LMA - IRAS correlation function. We have used
multiple cubes of the provided CDM space making use of the periodic boundary
conditions to simulate galaxies and X-ray sources up to v = 9000 and
20000 $km\,s^{-1}$ for the correlations with
the 2~Jy and 0.7~Jy samples respectively. The number density of the
provided CDM particles ($5.1 \times 10^{-3}\,h_{50}^{-3}\,Mpc^{-3}$)
is about the same as the number
density of the IRAS galaxies. We assigned each CDM particle an IRAS luminosity
drawn from the 2 Jy selection function (Strauss et al. 1992a) to simulate
the 2 Jy sample and used the luminosity
function by Saunders et al. (1990)  to simulate the 0.7 Jy sample.
We have also assigned a fraction of particles X-ray luminosities as
described in the following paragraphs. Then all-sky X-ray surface brightness
distributions contributed by these sources and flux limited samples of IRAS
galaxies were simulated and correlation coefficients were calculated
in $3\deg \times 3\deg$ square cells for a few hundred
times each model. Running a few hundred
Monte-Carlo simulations with the smearing process with the
A2 instrumental PSF in each run is computationally impractical.
Instead, we have
multiplied the $W_{xg}$ distribution simulated with the non-smeared square
cells by an attenuation factor. The factor should be the weighted mean of
$s_p$ and $s_c$ (\S~\ref{sec:ana}). The attenuation factor for each case was
dermined by the mean of five simulations where we evaluated the $W_{xg}$
values for both non-smeard and smeared cases.

We have included components [A] (AGNs with $L_{x,44}\gtrsim 0.01$)
and [B] lower X-ray luminosity sources defined in \S~\ref{sec:ana}
of X-ray populations in the simulations. In modeling component [A] in
the simulation, we used the following information: (1) the 60~$\mu m$
luminosity distribution (Saunders et al. 1990; Strauss et al. 1992a ),
(2) the X-ray luminosity distribution for AGNs with
$L_{x,44}\gtrsim 0.01$ (G92; Piccinotti et al.\ 1982), and (3) the
$L_x$ vs $L_{60}$ correlation for these AGNs (G92). The luminosity
correlation (3) was built into the simulation as follows. When an X-ray
luminosity $L_{x,44}(\geq 0.01)$ is assigned to a particle,
the corresponding IRAS luminosity was drawn from a
galaxy randomly chosen from the LMA sample galaxies  which
have X-ray luminosities between half and twice that of the assigned one
(see Fig.~\ref{fig:lma}).
 Component [B] is added for some models by randomly choosing a certain
fraction of CDM particles (which are not a part of [A]) and assigning
them a uniform X-ray luminosity with $L_{x,44}< 0.01$
(i.e., no luminosity correlation is assumed for compponent [B]).
We have assumed the same clustering property for the component
[B] sources.
We have considered several models to search for the
volume emissivity which fits the observed correlations well.

 The $\avrg{\delta I\, \delta N}$ values have been calculated
for 300 times each model, where $I$ and $N$ are the X-ray surface brightness
and the IRAS galaxy surface number density of the {\em simulated} particles.
These values are then divided by $\avrg{I_{CXB}}{N}$, where $\avrg{I_{CXB}}$
is the {\em real} mean CXB intensity, and corrected for the smearing effect
as discussed above. The distribution of the corrected  values can now be
directly compared with the observed $W_{xg}$ and its bootstrap
histogram. The simulation results are compared with the observations
and analytical predictions in \S~\ref{sec:comp}

\subsection{Comparison with the Observations} \label{sec:comp}

 Because the result from the 2~Jy 3500 - 8000 $km\,s^{-1}$ bin is least
sensitive to the local large scale structures (see Table~\ref{tbl:corr}
for the supergalactic latitude divisions) and also because the IRAS luminosity
functions from various works match very well with one another at higher
luminosities (e.g. Saunders et al. 1990; Strauss et al. 1992a), the most
reliable quantitative estimates for the local volume emissivity should be
drawn from the 2~Jy sample 3500-8000 $km\,s^{-1}$ bin. This subsample has
an advantage over the 0.7 Jy sample also because it has a well-defind
edges in the redshift space, although statistial errors are somewhat
larger.

 A model showing a good fit to the observed correlation for the
2~Jy 3500 - 8000 $km s^{-1}$ bin consists of only component [A] with
$\rho _{xA,38}=4.3$ (model I).
This value is within the error of the volume emissivity inferred from the
luminosity function by G92. Fig.~\ref{fig:hist} compares the
$W_{xg}$ histograms of model I simulation runs (300 for each; Thick
Solid Lines)
along with the observed values and their bootstrap histograms (Hatched).
Also shown for reference is the simulated histogram of another model
with both components with $\rho _{xA,38}=4.3$ and $\rho_{xB,38}=3.0$,
where 15\% of the CDM particles have $L_{x,44}=0.0039$ as
component [B] (model II; 100 runs each: Dashed lines).
Table~\ref{tbl:cmp} compares
the observed correlation (with the standard deviation of the bootstrap
histogram $\sigma_{bs}$) with the median, mean, and the standard deviation
($\sigma_{mc}$) of the Monte-Carlo runs for model I. The results of the
analytical calculations for model I is also shown. In any case, the
analytical values are somewhat larger than the simulated mean values.
The discrepancy is probably caused by the approximation in Eq.~\ref{eq:pcr2}
($|{\bf R_1} - {\bf R_2}| \ll R_1,\,R_2$).

Fig.~\ref{fig:hist} and Table~\ref{tbl:cmp} shows that model I predicts
somewhat larger correlations than observed for the 2~Jy 500-3500 $km s^{-1}$
division and the 0.7~Jy bin, but within 1 sigma errors. These two have
larger weight in the nearby universe and thus are more subject to the local
over/under densities and behavior of the IRAS luminosity function. In any
case, Fig.~\ref{fig:hist} shows that model II is certainly rejected.

It is noteworthy that the spread  of the $W_{xg}$ values from the
Monte-Carlo simulations for the model I are similar to
the bootstrap histogram for corresponding observations.
In particular, the  500 - 3500 $km\, s^{-1}$
division of the 2 Jy sample shows the wider spreads of $W_{xg}$ in both the
bootstrap and the Monte-Carlo histograms compared with those of the
3500 - 8000 $km\,s^{-1}$ division, although these two divisions contain
roughly the same number of galaxies. This means that the bootstrap
method is a good estimator of the main source of the spread,
i.e., the shot noise due to the finite number of X-ray sources contributing
to the correlation signal. The tendency that the simulation histogram has
a tail at higher $W_{xg}$ values can be understood in terms
of the rare occasions of a few extremely high flux sources contributing
to the correlation signal significantly.

Using the 2~Jy 3500-8000 sample vs CXB correlation as the optimum
estimator, the comparison of the simulation and observed correlation
leads to an estimated  volume emissivity of $\rho_{x,38}=4.3\pm1.2$
assuming that the IRAS galaxy vs X-ray emitter correlation can well be
represented by a  power law form of $\gamma=1.8$ and
$r_0=400\,km\,s^{-1}$. This is appropriate for the X-ray AGN - IRAS galaxy
correlation function (Fig.~\ref{fig:xi}). The error corresponds to
the 1$\sigma$ of the bootstrap runs of the observed correlation,
which is approximately equal to  the 1$\sigma$ of the simulated
correlations. Considering that
the AGN X-ray luminosity function of G92  (for $L_{x,44} \approx 0.01$)
gives $\rho_{x,38}=4.1_{-1.7}^{+1.3}$, the total volume emissivity
derived from the IRAS - CXB correlation allows the
local volume emissivity of the low luminosity objects of
$\rho_{xB,38}\lesssim 2$.

\section{Discussion}
\label{sec:dis}

%
%
%


 Fig.~\ref{fig:sel} shows that the clustering
term contribution  dominates the Poisson term (\S~\ref{sec:corr})
contribution to the correlation in our experiment.
It means that our results are insensitive to the details of the
$L_x - L_{60}$ correlation but sensitive to the clustering property between
IRAS galaxies and contributing X-ray sources. This is complementary to
the work by  Carrera et al. (1994) of the similar nature using the
GINGA scan data and the IRAS 0.7  Jy sample, where the Poisson term
contribution is dominant.
The factor of $\sim 2$ deficit compared
with the power-law used in the
X-ray AGN vs IRAS galaxy spatial correlation function (Fig.~\ref{fig:xi})
near $r\sim 150 km\,s^{-1}$ causes  $R_c$ to be reduced by 5 -- 9 \%,
allowing only 4 -- 7 \% more volume emissivity for the given correlation
signal.

The X-ray AGN vs IRAS
galaxy correlation function may not be a good description
of the clustering property between lower-luminosity X-ray sources
and IRAS galaxies.
If there are X-ray sources which are clustered with IRAS sources
only weakly, one may squeeze
more volume emissivity from these sources without
violating the observed correlation signal.
However, we can limit the contributions from lower X-ray luminosity
galaxies because  there is a reasonable range of spatial cross-correlation
functions among various galaxy catalogs (e.g. Lahav, Nemiroff, \& Piran 1990).
For example, using the IRAS auto-correlation parameters
($\gamma \approx 1.65$, $r_0 \approx 400 km\, s^{-1}$),
$R_c$ typically reduces
by $\sim$ 15 \% allowing slightly (by about 10 \%) more volume emissivity.
If we use the optical correlation parameters,
($\gamma \approx 1.8$, $r_0 \approx 500 km\, s^{-1}$), $R_c$ is
typically about 50 \% larger, allowing less volume emissivity.
Thus our result gives a strong constraint on the volume emissivity from the
low X-ray luminosity sources such as starburst galaxies and liners under the
reasonable assumption that their clustering property is not very
different from that of known galaxies.
The CDM model we have used in our
Monte-Carlo simulations well represents $\xi_{xg}(r)$ (Fig.~\ref{fig:xi})
at smaller scales ($r \lesssim 1000\,km\,s^{-1}$). However, it is known that
the observed large scale power of the galaxy distribution exceeds that
of the CDM model. If such large scale structures affect the correlation
signal, the estimated volume emissvity should be smaller, giving
a more strict upper limit to the contribution of the low lumninosity
objects to the local X-ray volume emissivity. A quantitative estimation
of the effect may be obtained by comparing the spherical harmonic
powers of the CXB surface brightness and the IRAS galaxies (c.f. Scharf et al.
1992); this may be addressed further in the future.

 If the case of no evolution and the energy spectral index of 0.7
appropriate for nearby AGNs in the $\Omega_0 = 1$ universe, the effective
look-back factor (Boldt 1987; Lahav et al. 1993) is $f=0.46$. The low local
volume emissivity derived with our experiment explains
about 20$\pm$ 5\% of the total 2 -- 10 keV CXB intensity under these
assumptions.
If these sources with the same spectrum have undergone a luminosity
evolution in 2 -- 10 keV similar to that in the soft X-ray band
(Boyle et al. 1993),
i.e. $L_x \propto (1+z)^{2.7}$ up to $z_{max}\sim 2$ for the
$\Omega_0 =1$ universe, the corresponding effective look-back factor
is $f=1.46$ and
explains about 65\% of the 2 -- 10 keV CXB. If we include sources with $z>2$,
the fraction becomes larger.  Although our results suggest that the
contributions of low luminosity sources such as star-forming galaxies
and liners to the 2 -- 10 keV volume emissivity {\em at present} is small,
our results do not exclude the possibility that such sources have
undergone stronger evolution (luminosity and/or number) contributing
significantly to the total CXB intensity in the past. Therefore it is
important to investigate the correlation property between the
CXB surface brightness and number counts from catalogs of higher
redshift objects.

The low volume emissivity implied by this work is hardly consistent
with the independent estimate of the local volume emissivity using the all-sky
X-ray dipole moment and the Local Group's peculiar motion
(JLMB92; Boldt 1990), i.e. $\rho_{x,38}\sim 30 (b_x\Omega _0^{-0.6})^{-1}$
predicting $b_x\Omega _0^{-0.6}\sim 7$.
Using the apparent dipole saturation at $v\approx 4500\,km\,s^{-1}$
(the dipole momemt of IRAS galaxies also seems to saturate at about the
same distance), Miyaji\& Boldt (1990); Miyaji, Jahoda, \& Boldt (1991) derived
$b_x\Omega _0 ^{-0.6} \approx 2.6\pm 0.5$ for X-ray selected AGNs. This kind
of estimation using flux-limited catalogs, however, could be subject to an
misestimation by a factor of 2 or more (e.g Strauss et al. 1992b; Peacock 1992
; Lahav, Kaiser, \& Hoffman 1990; Juszkiewics, Vittorio, \& Wyse 1990)
considering that the mass within the
apparent dipole saturation does not necessarily account for all the
gravitational acceleration at the local group, even though the flux dipole
within that depth appears to align with the peculiar velocity.
There still are uncertainties on the all-sky extragalactic X-ray dipole
in the subtraction of the galactic component and structures
behind the galactic plane. Apparently more study is needed to pursue this
comparison further.

\section{Conclusion}

The zero-lag cross correlation between the Cosmic X-ray Background and
the IRAS galaxy surface number density has been investigated. Two flux-limited
IRAS samples, i.e. the 2 Jy sample with redshift information and the 0.7 Jy
projected sample, are used to cross-correlate with the all-sky hard X-ray
map from the HEAO 1 A2 experiment. The cross-correlation study gives an
statistical estimation of the local volume emissivity from the faint X-ray
sources which are not resolved as point sources.

 We have detected the zero-lag correlation signals between the X-ray
surface brightness and the IRAS galaxy counts in the 9 $deg^2$ cells
of $W_{xg}\sim (3-11) \times 10^{-3}$ for selected IRAS samples. Both
Poisson and clustering effects contribute to the
correlation signal. The correlation between far infrared and X-ray
luminosities of galaxies affects the Poisson contribution of the
correlation. We have
developed an analytical formulation relating $W_{xg}$ and the X-ray volume
emissivity including these effects. We have also made Monte-Carlo simulations
using the particles from a CDM simulation, which fairly represent the
clustering properties of galaxies at the scales affecting our correlation
signal, by  assigning these particles X-ray and infrared luminosities
and observing  through the same cells as the real observations.

In the case of our  observation, the clustering term is the dominant
term of the correlation and thus the result is insensitive to the detail
of the far infrared - X-ray luminosity correlation. The volume emissivity
estimated from the correlation strength between the X-ray surface
brightness and IRAS 2~Jy 3500 - 8000 $km\,s^{-1}$ sample, which is
least subject to systematic errors in the IRAS luminosity function
and local large scale structures, is $\rho_{x,38}=4.3\pm1.2$.
This can be explained by AGNs with $L_{x,44}>0.01$ alone, whose volume
emissivity can be evaluated from the X-ray luminosity function.
Thus our correlation study implies that the contribution of lower luminosity
sources (e.g. star forming galaxies and liners) to the local volume emissivity
is not larger than that of AGNs and could be substantially smaller.
 The derived local volume emissivity (2 - 10 keV) explains about 20\%
and 65\% of the CXB intensity with no evolution and a luminosity evolution
(up to $z \sim 2$) similar to that of AGNs in soft X-rays
(Boyle et al. 1993) respectively.

\acknowledgments

We thank Simon White for providing us with his CDM simulation
and Bruce Grossan for sending us his Ph. D. thesis and allowing us to use
information from it. We also thank Francisco Carrera, Andy Fabian,
Xavier Barcons and Darryl Leiter for stimulating discussions. We also
thank the referee, Gianfranco DeZotti for helpful comments. This
paper is a part of the result of research toward TM's fulfillment
of the requirements of the Ph.D. degree at the University of Maryland.
This work is partially supported by an ADP grant to KJ.
TM appreciates the hospitality of  Institute of Astronomy, Cambridge
during his visit. OL thanks Laboratory for High Energy Astrophysics,
Goddard Space Flight Center for the hospitality during his visit.

\appendix
\section*{Appendix}
\section {The Poisson Term}
\label{app:po}
We here develop the formulation of the cross-correlation between the
intensity at a wavelength and a number density of objects selected
by the flux at another wavelength due to the Poisson process. Here
we use the example of X-ray intensity and IRAS galaxies. For
simplicity, we consider here observations through cells with
a square profile and express fluxes and number counts per cell by
script characters.

 Suppose we have two pupulations of objects with ${\cal N}_a$
and ${\cal N}_b$ per cell. The cross correlation of counts
at zero-lag square-profile cells of solid angle $\Omega$ is
separated into Poisson
and clustering terms:
\begin{equation}
\avrg{\delta {\cal N}_a\,\delta {\cal N}_b}
= \avrg{{\cal N}_o} + [clustering\,\, term],
\label{aeq:nab}
\end{equation}
where ${\cal N}_o$ is the number of objects per cell which overlap in both
populations $a$ and $b$ (Lahav 1992). Hereafter in this appendix,
we only consider
the Poisson term of the correlation. Suppose we are to correlate the
{\em flux} from population $a$ per cell, considering only the Poisson term
(${\cal I}_a$) with the number counts ${\cal N}_b$, then:
\begin{equation}
\avrg{\delta{\cal I}_a\, \delta{\cal N}_b}= \bar{f}_o \avrg{{\cal N}_o},
\label{aeq:fono}
\end{equation}
where $\bar{f}_o$ is the mean flux of the {\em overlapped} objects, which
is, in general, different from the mean flux of the population $a$ objects.

As a simple illustration of the effect of the luminosity
correlation, let us first consider the case where we have an
X-ray emitting population with a luminosity function $\Phi_x(L_x)$
(normalized to the spatial number density). Let us cross-correlate the X-ray
fluxes ${\cal I}_x$ and number counts ${\cal N}_x$ of an X-ray flux limited
(at $f_{x,lim}$) sample:

First, let us consider the contribution of the objects in a thin shell
at a distance $R$ from us
$[R,R+\Delta R]$. The number count from the shell ($\Delta_R {\cal N}_x$)
is:
\begin{equation}
\Delta_R {\cal N}_x
=\int_{4\pi R^2 f_{x,lim}}^{\infty}dL_x\Phi_x(L_{x}) R^2 \Omega
\Delta R,
\end{equation}
and the mean flux of the sources in the flux limited sample from
the thin shell (corresponding to the {\em overlapped} objects in
Eq.~\ref{aeq:fono}):
\begin{equation}
\bar{f}_x(R) =  (\Delta_R {\cal N}_x)^{-1}
\int_{4\pi R^2 f_{x,lim}}^{\infty}dL_x \frac{L_x}{4 \pi R^2}
\Phi_x(L_{x}) R^2 \Omega \Delta R .
\label{aeq:fxbar}
\end{equation}
Note that this is a function of $R$.
Thus the correlation $\avrg{\delta{\cal I}_x\,\delta{\cal N}_x}$ from the
sources between distances $[R_{min},R_{max}]$ can be found by radially
integrating the contributions from radial shells:
\begin{equation}
\sum _{shells} \bar{f}_x(R) \Delta_R {\cal N}_x =
\int_{R_{min}}^{R_{max}} dR
\left[ \frac{1}{4\pi}\int_{4\pi R^2 f_{x,lim}}^{\infty}dL_x\, L_x
\Phi_x(L_{x})\Omega \right].
\end{equation}
Using an expression of total X-ray volume emissivity $\rho_{xp}$ from the
sources and the definition
of the X-ray selection function $P_x(R)$:
\begin{mathletters}
\begin{equation}
\rho_{xp} = \int_0^{\infty} dL_x L_x \Phi_x(L_x),
\end{equation}
\begin{equation}
P_x(R)\equiv \frac{\int_{4\pi R^2 f_{x,lim}}^{\infty} dL_x\, L_x \Phi_x(L_x)}
{\int_{0}^{\infty} dL_x\, L_x \Phi_x(L_x)},
\end{equation}
\end{mathletters}
emphasizing that $P_x(R)$ is the X-ray luminosity weighted
radial selection function, we get,
\begin{equation}
\avrg{\delta {\cal I}_x\,\delta {\cal N}_x}=\frac{\rho_x \Omega}{4\pi}
\int_{R_{min}}^{R_{max}} dR\, P_x(R).
\label{aeq:ixnx}
\end{equation}

When we have a galaxy sample which is flux limited at a different
wavelength ($60 \mu m$ in this case, with the luminosity function
$\Phi_{60}(L_{60})$ and the limiting flux of $f_{60,lim}$) to
correlate with the X-ray surface brightness, we have
to modify the expression of $P_x(R)$ accordingly. In this case, the
contribution of the thin shell to
$\avrg{\delta{\cal I}_x\,\delta{\cal N}_{60}}$ is
$\bar{f}_x(R) \Delta_R {\cal N}_{60}$, where $\bar{f}_x(R)$ is now the mean
X-ray flux of the objects which are in the $60 \mu m$ flux limited sample
at the distance $R$, i.e. the objects with $L_{60}\geq 4\pi R^2 f_{60,lim}$.
Then, this can be expressed as:
\begin{eqnarray}
\bar{f}_x(R) \Delta_R {\cal N}_{60}
=\int_{4\pi R^2 f_{60,lim}}^{\infty}dL_{60}\Phi_{60}(L_{60})
\int_0^{\infty} dL_x \frac{L_x}{4\pi R^2} p(L_x|L_{60}) R^2 \Omega \Delta R
\nonumber \\
\equiv \frac{\Omega}{4\pi}
\int_{4\pi R^2 f_{60,lim}}^{\infty}dL_{60}\bar{L_x}(L_{60})\Phi_{60}(L_{60})
\Delta R,
\end{eqnarray}
where $p(L_x|L_{60})dL_x$ is the normalized probability that an
object has an X-ray
luminosity between $Lx$ and $L_x+dL_x$ given that its $60 \mu m$ luminosity
is $L_{60}$ and $\bar{L_x}(L_{60})$ is the mean X-ray luminosity of objects
with $60 \mu m$ luminosity of $L_{60}$, i.e.
$\bar{L_x}(L_{60})=\int_0^{\infty}dL_x\,L_x\,p(L_x|L_{60})$. As before, we can
radially integrate this to find ($I_x$ \& $N_{60}$ are defined as quantities
{\em per solid angle}):
\begin{equation}
W_{xg}\avrg{I_x}\avrg{N_{60}}\equiv \avrg{\delta I_x\,\delta N_{60}}=
\frac{\avrg{\delta{\cal I}_x\,\delta{\cal N}_{60}}}{\Omega^2}
= \frac{\rho_{xp}}{4 \pi \Omega} \int_{R_{min}}^{R_{max}} dR\,P_x (R)
\label{aeq:wxgp}
\end{equation}
\begin{equation}
P_x(R)=
\frac{\int_{4\pi R^2
f_{60,lim}}^{\infty}dL_{60}\bar{L_x}(L_{60})\Phi_{60}(L_{60})}
{\int_0^{\infty}dL_{60}\bar{L_x}(L_{60})\Phi_{60}(L_{60})},
\label{aeq:pxr1}
\end{equation}
with a note that the denominator of $P_x(R)$ is equal to $\rho_{xp}$. This
is the square profile cell case of Eqs.~\ref{eq:etap} - \ref{eq:pxr}.

As a preparation to practically evaluate $P_x(R)$ from available
information, let us consider the bivariate function
(c.f. Sodr\'e \& Lahav 1993)
in the $(L_x, L_{60})$ space $\Psi (L_x,L_{60})dL_xdL_{60}$, defined as the
mean space density of galaxies within the luminosity-luminosity space
element $dL_xdL_{60}$.
 From the Bayes theorem,
$\Psi (L_x,L_{60})= p(L_x|L_{60}) \Phi_{60}(L_{60})$.
Then the numerator
of the expression of $P_x(R)$ in Eq.~\ref{aeq:pxr1} is:
\begin{equation}
\int_{4\pi R^2 f_{60,lim}}^{\infty}dL_{60}\bar{L_x}(L_{60})\Phi_{60}(L_{60})
=\int _{4\pi R^2 f_{60,lim}}^{\infty}dL_{60}\int_0^{\infty}dL_x\,L_x
\Psi (L_x,L_{60}).
\label{aeq:pxrbv}
\end{equation}
Using the available sample of X-ray flux selected galaxies with measured
redshifts and $60 \mu m$ fluxes (the LMA sample [G92] in this work),
the bivariate function
$\Psi (L_x,L_{60})$ can be constructed above some minimum X-ray
luminosity  $L_{x,min}$ (defined by the sample). Neglecting the
clustering effect, the bivariate function can be estimated by plotting
the objects in the sample ($L_{x,i},\,L_{60,i}$) on the luminosity-
luminosity plane weighted by $V_{max}(L_{x,i})^{-1}$,
where $V_{max}(L_x)$ is the maximum
volume of space where an object with $L_x$ would be in the sample.
Dividing by  $V_{max}(L_x)$ compensates for the effect of using the
X-ray flux-limited sample and also gives the proper normalization.
Then the
double integral (lower integration limit for $L_x$ is now $L_{x,min}$ instead
of zero) in Eq.~\ref{aeq:pxrbv} can then be expressed by the sum
over the sample with $L_{60}\geq 4\pi R^2 f_{60,lim}$:
\begin{equation}
\int _{4\pi R^2 f_{60,lim}}^{\infty}dL_{60}\int_{L_{x,min}}^{\infty}dL_x\,L_x
\Psi (L_x,L_{60}) \sim \sum_{L_{60,i}\geq 4\pi R^2 f_{60,lim}}
\frac{L_{x,i}}{V_{max}(L_{x,i})}.
\end{equation}
This is Eq.~\ref{eq:pxr2} in \S~\ref{sec:ana}
Because of the lower X-ray luminosity limit from the available sample,
we emphasize that the above expression actually represents the portion
of $\rho_{xp}P_x(R)$ contributed by the sources with $L_x\geq L_{x,min}$,
expressed in terms of $\rho_{xpA}$,$P_{xA}(R)$, and $R_{pA}$ in the main
text.

\section{Effect of the Non-Square Profile of the Effective Cell on the
Poisson Term}
\label{app:beam}
We use the same notations as in \S~\ref{sec:ana} in the main text and
appendix A  unless otherwise noted.
In many cases (including this work), we measure the X-ray
intensities in observing cells with a size comparable to the
instrumental point spread function (PSF). In our case, the IRAS
galaxy distributions have also been smeared with the same PSF
to perform the correlation. In that case,
we have to consider the effective cell profile
$B_{ec}(\hat{R}-\hat{R}_0)$, which is the convolution of the
PSF $B_{psf}(\hat{R}-\hat{R}_0)$ with the square cell profile
$B_{sq}(\hat{R}-\hat{R}_0)$(=1 in the cell; =0 outside of the cell),
\begin{equation}
B_{ec}(\hat{R}-\hat{R}_0) \propto \int d\Omega_1
B_{psf}(\hat{R}-\hat{R}_1)\, B_{sq}(\hat{R}_1-\hat{R}_0),
\end{equation}
normalized to unity at the center.

Now let us consider the quantities observed through this effective
cell (expressing by script characters as before).
Then, noting that $N$ and $I$ (surface number density and surface
brightness) are the functions of the sky position,
\begin{mathletters}
\begin{eqnarray}
\avrg{\cal N}=\int d\Omega N(\hat{R}) B_{ec}(\hat{R}-\hat{R}_0) =
\avrg{N}\Omega_{ec}, \\
\avrg{\cal I}=\int d\Omega I(\hat{R}) B_{ec}(\hat{R}-\hat{R}_0) =
\avrg{I}\Omega_{ec}
\end{eqnarray}
\end{mathletters}
In the Poisson term of the correlation, since both flux and surface
number density  of the {\em overlapped} objects (see Eq.~\ref{aeq:fono})
should be weighted by
the profile, noting that $N_o(\hat{R})$ is the sum of randomly
placed $\avrg{N_o}$ delta functions per solid angle:
\begin{equation}
\avrg{\delta{\cal I}\,\delta{\cal N}}=\int d\Omega \bar{f}_o N_o(\hat{R})
B_{ec}^2(\hat{R}-\hat{R}_0)
=\bar{f_x}\avrg{N}\int d\Omega B_{ec}^2(\hat{R}-\hat{R}_0).
\end{equation}
Therefore, reading $\bar{f}_o$ as $\bar{f}_x$ and $N_o$ as $N$,
\begin{equation}
W_{xg}\avrg{I}\avrg{N}=\frac{\int d\Omega B_{ec}^2(\hat{R}-\hat{R}_0)}
{\Omega_{ec}^2} \bar{f_x}\avrg{N}.
\label{eq:wxgap}
\end{equation}

 Thus the PSF smearing effect, or more in general, the case of non-square
profile of effective cells, can be taken into account by replacing
$\frac{1}{\Omega}$ in Eq.~\ref{aeq:wxgp} by
$\int d\Omega B_{ec}^2(\hat{R}-\hat{R}_0) \Omega_{ec}^{-2}$
( Eq.~\ref{eq:etap}).

\section{The Clustering Term with a Power-Law Correlation Function}
\label{app:plcl}
In the case of a power law spatial correlation function
$\xi _{xg} = (r/r_{0})^{-\gamma}$, the
clustering term $\hat{\eta}_c$ (Eq.~\ref{eq:etac})
can be expressed in a more convenient form under the approximation that
the clustering scale length is much smaller than the distance to the
objects ($|{\bf R_1}-{\bf R_2}|\ll R_1,\, R_2$). Under this approximation,
changing the variables to $u=R_1 - R_2$,$\, x=\frac{R_1 + R_2}{2}$, then
$R_1 \approx x$ and
$r\equiv|{\bf R_1}-{\bf R_2}| \approx (u^2 + x^2 \theta ^2)
^{\frac{1}{2}}$,
where $\theta$ is the angle between ${\bf R_1}$ and  ${\bf R_2}$.
Then Eq.~\ref{eq:etac} can be rewritten as:
\begin{equation}
\hat{\eta}_c \approx \frac{\avrg{n}\rho_x\,r_0^{\gamma}}{\Omega_{ec}^2}
\int dx\,x^2P(x) \int \int d\Omega_1\,d\Omega_2
B_{ec}(\hat{R}_1-\hat{R_0})\,B_{ec}(\hat{R}_2-\hat{R}_0)
\int_{-\infty}^{\infty}du\,(u^2+x^2\theta^2)^{-\frac{\gamma}{2}}.
\end{equation}

The integration in $u$ can be calculated (Peebles 1980, Eq. 52.9):
\begin{equation}
\int_{-\infty}^{\infty}du\,(u^2+x^2\theta^2)^{-\frac{\gamma}{2}}
= H_{\gamma}\,(x\theta)^{1-\gamma},\,\,\, H_{\gamma}=
\frac{\Gamma(\frac{1}{2})\,\Gamma(\frac{\gamma-1}{2})}
{\Gamma(\frac{\gamma}{2})}.
\end{equation}
For example, $H_{1.8}=3.68$ and $H_{1.65}=4.29$.

The double integral over the solid angles:
\begin{equation}
X = \int \int d\Omega_1\,d\Omega_2\,\theta^{1-\gamma}
B_{ec}(\hat{R}_1-\hat{R}_0)\,B_{ec}(\hat{R}_2-\hat{R}_0)
\label{aeq:intomega}
\end{equation}
 is now separated from the radial
integrations. This can be integrated numerically by the Monte-Carlo method.
For the $a \times a$ square cell case instead of the real effective cell,
this double integral is expressed as $C_{\gamma}a^{5-\gamma}$
 with $C_{1.8}=2.25$ and
$C_{1.65}=1.87$ (Totsuji \& Kihara 1969).
Then defining $A_{\gamma} = H_{\gamma} C_{\gamma}$ immediately gives
Eq.\ref{eq:pcr2}.
For our convolved effective cell,
$X$ is $\approx 99\, deg^{3.2}$ and
$\approx 102\, deg^{3.35}$ corresponding
to $s_c=$ 0.74 and 0.77  (second raw of Eq.~\ref{eq:etac})
for $\gamma=$ 1.8 and 1.65 respectively.

\begin{table*}
\begin{center}
\caption{The IRAS Galaxy -- CXB Correlation: Results}
\vspace{.25in}
\begin{tabular}{crrcrccc}
\tableline
IRAS & $v_{min}$ & $v_{max}$ & $|SGB|$ & $N_{cells}$
& $\avrg{I}$ & $\avrg{N}$ & $W_{xg}$ \\
Sample & \multicolumn{2}{c}{[$km\,s^{-1}$]}
& [deg] & $\ldots$ & [$cts\,s^{-1}\,deg^{-2}$] & [$deg^{-2}$]
& $\ldots$ \\
\tableline
2~Jy &500 &8000 & $\ldots$ &2328 & 0.71 & $5.4\times 10^{-2}$ &
$(7.6\pm 1.6)\times 10^{-3}$\\
2~Jy &500 &8000 & $\geq 20\deg$ & 1413 & 0.71 &
$4.5\times 10^{-2}$ & $(5.6\pm 1.9)\times 10^{-3}$\\
2~Jy &500 &8000 & $<20\deg$ & 915 & 0.71 &
$6.7\times 10^{-2}$ & $(9.3\pm 2.6)\times 10^{-3}$\\
2~Jy\tablenotemark{a} &500 &8000 & $\ldots$ & 2275 & 0.71 &
$5.2\times 10^{-2}$ & $(4.1\pm 1.2)\times 10^{-3}$\\
\\
2~Jy & 500 & 3500 & $\ldots$ & 2328 & 0.71 &
$2.8\times 10^{-2}$ & $(8.6\pm 2.8)\times 10^{-3}$\\
2~Jy & 500 & 3500 & $\geq 20\deg$ & 1413 & 0.71 &
$2.2\times 10^{-2}$ & $(4.3\pm 2.8)\times 10^{-3}$\\
2~Jy & 500 & 3500 & $<20\deg$ & 915 & 0.71 &
$3.8\times 10^{-2}$ & $(11.3\pm 4.3)\times 10^{-3}$\\
\\
2~Jy & 3500 & 8000 & $\ldots$ & 2328 & 0.71 &
$2.5\times 10^{-2}$ & $(6.3\pm 1.7)\times 10^{-3}$\\
2~Jy & 3500 & 8000 & $\geq 20\deg$ & 1413 & 0.71 &
$2.3\times 10^{-2}$ & $(5.3\pm 2.0)\times 10^{-3}$\\
2~Jy & 3500 & 8000 & $<20\deg$ & 915 & 0.71 &
$2.9\times 10^{-2}$ & $(6.9\pm 2.8)\times 10^{-3}$\\
\\
0.7~Jy & \multicolumn{2}{c}{$\ldots$} & $\ldots$ & 2379 & 0.71 &
$2.7\times 10^{-1}$ & $(3.6\pm 0.7)\times 10^{-3}$\\
0.7~Jy & \multicolumn{2}{c}{$\ldots$} & $\geq 20\deg$ & 1453 & 0.71 &
$2.6\times 10^{-1}$ & $(2.8\pm 0.8)\times 10^{-3}$\\
0.7~Jy & \multicolumn{2}{c}{$\ldots$} & $< 20\deg$ & 926 & 0.71 &
$2.9\times 10^{-1}$ & $(4.5\pm 1.0)\times 10^{-3}$\\
\tableline
\label{tbl:corr}
\end{tabular}
\end{center}
\tablenotetext{a}{All AGNs in Piccinotti et al. (1982) are excluded.}
\end{table*}
\clearpage
\begin{table*}
\begin{center}
\caption{Comparison of Observations with Model I ($\rho_{x,38}=4.3$)}
\vspace{.25in}
\begin{tabular}{crrrcccccc}
\tableline
 & & & & \multicolumn{6}{c}{$W_{xg} \times 10^3$}\\
\multicolumn{2}{c}{IRAS Sample}
 & $R_{pA}$\tablenotemark{a} & $R_c$\tablenotemark{a} &
\multicolumn{2}{c}{observed} &
\multicolumn{3}{c}{simulated} & analytical\\
 & $[km\,s^{-1}]$ & \multicolumn{2}{c}{$[km\,s^{-1}]$}
 & nominal & $\sigma_{bs}$\tablenotemark{b} & median & mean
& $\sigma_{mc}$\tablenotemark{c} & \\
\tableline
2~Jy & 500-3500 & 2390 & 2560 & 8.6 & 2.8 & 10.7 & 13.0 & 6.1 & 17.  \\
2~Jy & 3500-8000 &  980 &  1100 & 6.3 & 1.7 & 6.2 & 6.4 & 1.9 & 7.9 \\
0.7~Jy &$\ldots$ & 6700 & 11000 & 3.6 & 0.7 & 4.7 & 5.0 & 1.5 & 6.5 \\
\tableline
\label{tbl:cmp}
\end{tabular}
\end{center}
\tablenotetext{a}{Effective depths defined in \S\S~\ref{sec:ana} \&
\ref{sec:appl}}
\tablenotetext{b}{The standard deviation of $W_{xg}$ for the bootstrap runs.}
\tablenotetext{c}{The standard deviation of $W_{xg}$ for the Monte-Carlo runs.}
\end{table*}


%
%
\clearpage
\begin{figure}
\caption{The $L_x$ vs $L_{60}$ Plot of X-ray AGNs}
The IRAS 60 $\mu m$ luminosity is plotted against the 2 -- 10 keV
X-ray luminosity for the LMA sample AGNs (G92) with v$\leq$
20000 $km\,s^{-1}$.
The 60 $\mu m$ luminosity is in the solar unit and the X-ray luminosity
is in $10^{44}h_{50}^{-2}erg\,s^{-1}$.
\label{fig:lma}
\end{figure}
\begin{figure}
\caption{The X-ray AGN - IRAS galaxy Correlation Function}
\label{fig:xi}
 The spatial correlation function $\xi_{xg}(r)$ between LMA AGNs (G92)
and IRAS 2 Jy galaxies is shown by filled hexagons with error bars. The
power-law function with $\gamma=1.8$, $\,r_0=400\,km\,s^{-1}$ is shown
by a dashed line. Open triangles show the spatial correlation of the
CDM particles (\S~\ref{sec:sim}) rescaled to the correlation length of
 $\,r_0=400\,km\,s^{-1}$.
\end{figure}
\begin{figure}
\caption{Comparison of Radial Selection Functions.}
The selection functions multiplied by the PSF smearing factors are drawn for
(a) the 2 Jy sample, and (b) the 0.7 Jy sample. These curves show
the relative contributions of the clustering and Poisson terms as
functions of $R$.
Solid:$s_pP(R)$, the IRAS 2 Jy and 0.7 Jy selection functions multiplied
by the Poisson smearing factor for reference;
Dashed:$s_cP_c(R)$, the effective radial selection
function for the clustering term defined in Eq.~\ref{eq:pcr2} multiplied
by the clustering smearing factor, and
Dot-dashed:$s_pP_x(R)$, the X-ray selection
function of the 2 Jy and 0.7 Jy samples evaluated using Eqs.~\ref{eq:pxr} \&
\ref{eq:pxr2} multiplied by the Poisson smearing factor.
The wiggle in the $P_x(R)$ curve is caused by the
discreetness of the summation in Eq.~\ref{eq:pxr2}.
\label{fig:sel}
\end{figure}
\begin{figure}
\caption{The Observed and Simulated Correlation Coefficients}
The observed correlation coefficients (arrows with the bootstrap
histograms) are compared with the Monte-Carlo simulations for two
models (model I: component [A] only ($\rho_{x,38}=4.3$) and model II:
including components [A] and [B] ($\rho_{x,38}=7.3$). The IRAS samples
used are: (a) the 2Jy sample with $500\leq V[km\,s^{-1}]<3500$,
(b) the 2 Jy sample with $500\leq V[km\,s^{-1}]<3500$, and
(c) the 0.7 Jy smaple. The histograms
in each figure are normalized to have the same area.
\label{fig:hist}
\end{figure}
%


\clearpage
\begin{references}
\reference Boldt, E. 1987, Phys.\ Reports, 146, 215
\reference ---., 1990, in IAU Colloquim 123, Observatories in Earth Orbit
and Beyond, ed. Y. Kondo (Dordrecht:Kluwer),451
\reference Boyle, B. J., Griffiths, R. E., Shanks, T., Stewart, G. C,
\& Georgantopoulos, I. 1993, \mnras, 260, 49
\reference Carrera, F. J. et al. 1994, in preparation
\reference Danese, L. et al. 1993, \apj, 412, 56
\reference Edge A. C., Stewart, G. C., Fabian, A. C., \& Arnaud, K. A.,
1990, \mnras, 245, 559
\reference Fabian, A. C. \& Barcons, X. 1992, \araa, 30, 429
\reference Georgantopoulos, I. \& Shanks, T. 1993, preprint
\reference Grossan, A. G. 1992, Ph. D. thesis, Massachusetts Institute of
Technology (G92)
\reference Griffths, R. E. \& Padovani, P. 1990, \apj, 360, 483
\reference Hasinger, G., Burg, R., Giacconi, R., Hartner, G., Schmidt, M.,
          Tr\"umper, J, \& Zamorani, G. 1993, \aap, 275,1
\reference Hasinger, G., Schmidt, M. \& Tr\"umper, J. 1991, \aap, 246, L2
\reference Huchra, J. \& Burg, G. 1992, \apj, 393,90
\reference Jahoda, K., Lahav, O., Mushotzky, R. F., \& Boldt, E. A. 1991,
           \apjl,378,L37 (JLMB91)
\reference ---. 1992, \apjl, 399, L107 (Erratum, JLMB92)
\reference Jahoda, K. \& Mushotzky, R. F. 1989, \apj, 346, 638
\reference Juszkiewics, R., Vittorio, N., \& Wyse, R. F. G. 1990,
         \apj, 349, 408
\reference Kogut, A. et al. 1993, \apj, 419,1
\reference Lahav, O. 1992,  in The X-ray Background, eds. Barcons, X.
   \& Fabian, A. C., p102, (Cambridge:Cambridge Univ. Press)
\reference Lahav, O. et al.\, 1993, Nature, 364, 693
\reference Lahav, O., Kaiser, N., \& Hoffman, Y. 1990, \apj, 352, 448
\reference Lahav, O., Nemiroff, J. R., Piran, T. 1990, \apj, 350, 119
\reference Leiter, D. \& Boldt, E. 1993, in preparation
\reference Meurs, E. J. A. \& Harmon, R. T. 1989, \aap, 206, 53
\reference Miyaji, T. \& Boldt, E. 1990, \apjl, 353, L3
\reference Miyaji, T., Jahoda, K.,\& Boldt, E., 1991, in After the First Three
          Minutes, AIP Conference Proceedings 222, eds. Holt, S. S., Bennett,
          C. L., \& Trimble, V., 431 (New York:AIP)
\reference Peacock, J. A. 1992, \mnras, 258, 581
\reference Peebles, P. J. E. 1980, The Large Scale Structure of the Universe,
           (Princeton:Princeton)
\reference Persic, M., De Zotti, G., Danese, L., Palumbo, G. G. C.,
Franceschini, A., Boldt, E. A., \& Marshall, F. E. 1989, ApJ, 344, 125
\reference Piccinotti, G. et al.\ 1982, \apj, 253, 485
\reference Remillard, R. A., 1994, preprint
\reference Rothschild R. et al.\ 1979, Sp.\ Sci.\ Instr., 4, 269
\reference Saunders, W., Rowan-Robinson, M., \& Lawrence, A. 1992, \mnras,
         258,134
\reference Saunders, W., Rowan-Robinson, M., Lawrence, A., Efstathiou, G.,
Kaiser, N., Ellis, R. S., \& Frenk, C. S., 1990, \mnras, 242, 318
\reference Shafer, R. A. 1983, Ph. D. Thesis, University of Maryland
\reference Shanks T., Georgantopoulos, I, Stewart, G.C, Pounds, K. A.,
    Boyle, B. J. \& Griffiths, R. E. 1991,  Nature, 353, 315
\reference Scharf, C. A., Hoffman, Y., Lahav, O., \& Lynden-Bell, D. 1992,
\mnras, 256, 229
\reference Sodr\'e, L., Jr. \& Lahav, O. 1993, \mnras, 260, 285
\reference Strauss, M. A., Davis, M., Yahil, A., \& Huchra, J. P. 1990,
    \apj, 361, 49
\reference ---. 1992a, \apj, 385, 421
\reference Strauss, M. A., Yahil, A., Davis, M., Huchra, J. P., \& Fisher, K.
        1992b, \apj, 397, 395
\reference Totsuji, H. \& Kihara T. 1969, \pasj, 21, 221
\reference Turner, E. L. \& Geller, M. J. 1980, \apj, 236, 1
\reference White, S. D. M. 1993, private communication
\end{references}
\end{document}